\documentclass[trackchanges,twocolumn]{aastex701}

\begin{document}

\title{JWST-DECO: The Impact of Accretion on Mid-Infrared Observable Water in Planet-forming Disks}

\author[0000-0002-0150-0125]{Jenny K. Calahan}
\affiliation{Center for Astrophysics \textbar\ Harvard \& Smithsonian, 60 Garden St., Cambridge, MA 02138, USA }
\email{jenny.calahan@cfa.harvard.edu}

\author[]{Tarisai Dziire}
\affiliation{New York University Shanghai, 567 West Yangsi Rd, Pudong, Shanghai, China, 200124}
\email{tid2005@nyu.edu}

\author[0000-0001-8798-1347]{Karin \"{O}berg}
\affiliation{Center for Astrophysics \textbar\ Harvard \& Smithsonian, 60 Garden St., Cambridge, MA 02138, USA }
\email{koberg@cfa.harvard.com}

\author[0000-0003-4335-0900]{Andrea Banzatti}
\affiliation{Department of Physics, Texas State University, 749 N Comanche Street, TX 78666}
\email{banzatti@txstate.edu}

\author[0000-0003-2076-8001]{L. Ilsedore Cleeves}
\affiliation{Department of Astronomy, University of Virginia, Charlottesville, VA 22904, USA}
\email{lic3f@virginia.edu}

\author[0000-0002-2692-7862]{Felipe Alarc\'{o}n}
\affiliation{Dipartimento di Fisica, Università degli Studi di Milano, Via Celoria 16, 20133 Milano, Italy}
\email{felipe.alarcon@unimi.it}

\author[0000-0003-0787-1610]{Geoffrey A. Blake}
\affiliation{Division of Geological \& Planetary Sciences, California Institute of Technology, Pasadena, CA 91125, USA}
\email{gab@gps.caltech.edu}

\author[orcid=0000-0002-5296-6232,gname='Maria Jose']{Mar\'ia Jos\'e Colmenares}
\affiliation{Department of Astronomy, University of Michigan, 1085 South University Avenue, Ann Arbor, MI 48109, USA}
\email{mjcolmen@umich.edu}  

\author[0000-0003-4907-189X]{Camilo Gonz\'alez-Ruilova}
\affiliation{Departamento de F\'isica, Universidad de Santiago de Chile, Av. Victor Jara 3659, Santiago, Chile}
\affiliation{Millennium Nucleus on Young Exoplanets and their Moons (YEMS), Chile}
\affiliation{Center for Interdisciplinary Research in Astrophysics and Space Science (CIRAS), Universidad de Santiago, Chile}
\email{camilo.gonzalez.ru@usach.cl}

\author[0000-0002-9959-1933]{Aashish Gupta}
\affiliation{Department of Astronomy, University of Virginia, Charlottesville, VA 22904, USA}
\email{aashishgupta@virginia.edu}

\author[0009-0009-2320-7243]{Claudio Hern\'{a}ndez-Vera}
\affiliation{European Southern Observatory, Alonso de Córdova 3107, Casilla 19001, Vitacura, Santiago, Chile}
\affiliation{Millennium Nucleus on Young Exoplanets and their Moons (YEMS), Chile}
\email{Claudio.HernandezVera@eso.org}

\author[0000-0001-8240-978X]{Till Kaeufer}
\affiliation{Department of Physics and Astronomy, University of Exeter, Exeter EX4 4QL, UK}
\email{T.Kaeufer@exeter.ac.uk}

\author[0000-0002-3291-6887]{Sebastiaan Krijt}
\affiliation{Department of Physics and Astronomy, University of Exeter, Exeter EX4 4QL, UK}
\email{s.krijt@exeter.ac.uk}

\author[0000-0003-1413-1776]{Charles J.\ Law}
\altaffiliation{NASA Hubble Fellowship Program Sagan Fellow}
\affiliation{Department of Astronomy, University of Virginia, Charlottesville, VA 22904, USA}
\email{cjl8rd@virginia.edu}

\author[0000-0003-0448-6354]{Tamara Molyarova}
\affiliation{School of Physics and Astronomy, University of Leeds, Leeds, LS2 9JT, UK}
\email{t.molyarova@leeds.ac.uk}

\author[0009-0008-8176-1974]{Joe Williams}
\affiliation{Department of Physics and Astronomy, University of Exeter, Exeter EX4 4QL, UK}
\email{jw1436@exeter.ac.uk}

\begin{abstract}
The inner few au of a protoplanetary disk hosts the majority of observed exoplanets and is the primary planet-forming zone of the disk. The mid-IR spectra of disks, with its rich forest of water lines, provides key insights into the composition of forming planets. One of the strongest trends seen with data from \textit{Spitzer} and now JWST is a correlation between the increase in water line flux and accretion luminosity of a system. We set out to reproduce and understand this trend by adding an accretion module to the thermo-chemical code DALI, and explore how viscous accretion heating and the addition of accretion luminosity impacts the 2D temperature structure and the observable water reservoir. We reproduce the trend that the observed water mass increases with accretion rate, with hot, warm, and cool water being more to less strongly correlated, respectively. Our model suggests that these trends are due to an increased emitting area with accretion rate, with some of the cool and warm population becoming hidden underneath an optically thick dust surface and being constrained to a smaller disk volume. This trend is driven by the accretion-related increase in central luminosity, while viscous heating centralized to the midplane has no impact on observed water mass. %With our fiducial thermo-chemical model, we also reproduce the observed water mass as retrieved by slab models. 
This analysis provides the necessary framework for future interpretation of JWST spectral features in the context of disk, stellar, and environmental properties. 
\end{abstract}

\section{Introduction}%\label{sec:Intro}
Protoplanetary disks contain the building blocks to form planets. Most known planets are found within the first few au of their host star \citep{Zhu21, Chambers01,Ogihara18}. This region of the protoplanetary disk is warm, between $\sim$100 to 1000~K, emitting at infrared wavelengths and is the expected site of the formation of potentially habitable planets. The space-based telescopes \textit{Spitzer} and JWST have been the flagship mid-infrared (IR) probes that have now have observed hundreds of protoplanetary disks. Although there is a wide diversity of signatures when looking at individual disks, some trends stand out that require explanation. Understanding the physical origin of these trends is essential for linking observed disk properties to their planet-formation potential. There are trends with stellar type, with more massive stars having less-line rich spectra than T Tauri systems \citep{Pontoppidan10,Salyk11,Banzatti17,Arulanantham25,Henning24, Xie23}. The lowest mass stars seem to display brighter hydrocarbon-rich emission relative to higher mass counterparts \citep{Pascucci13,Tabone23,Arabhavi25,Grant25}. Additionally, the flux ratio of hot/cold water lines appear to correlate with disk size or substructure \citep{Banzatti23,Banzatti20,RomeroMirza24, Gasman25,Krijt25}. However, one of the strongest trends that have been seen since \textit{Spitzer}, has yet to be critically explored: the correlation between line flux and accretion luminosity/rate \citep{Salyk11,Banzatti23,Banzatti25}.  

%forest of water lines\citep{Salyk08,Pontoppidan10,Banzatti25}, while others are extremely water-poor but hydrocarbon .  One of the clearest trends that has been seen since \textit{Spitzer}, is the trend of molecular line luminosities and accretion luminosity \citep{Salyk11,Banzatti23}.  

Accretion from the protoplanetary disk onto the star is a critical mechanism that drives disk evolution and is a primary way to disperse the disk \citep{Gorti16}. The transport of material inwards towards the star provides heat to the system, particularly in regions where accretion is most efficient \citep{Wang25}. A common initial assumption is that accretion primarily resides in the disk midplane, where the density of the disk is highest \citep[i.e.][]{Shakura1973,Lynden-Bell74,Hartmann98}. %Another possibility is that there is accretion of disk material in the atmosphere, where the disk has magnetorotational instability (MRI) due to low densities, thus higher rate of ionization, coupling atoms and molecules to the magnetic field, driving a loss of angular momentum \citep{Perez-becker11,Simon13}. Under either scenario, 
Material being deposited from the disk onto the star will additionally contribute to the total stellar luminosity, and that material can be approximated as a blackbody at $\approx$8,000~K \citep{McClure13}. Both the added luminosity due to accretion and the viscous accretion heating within the disk itself should increase the temperature of the disk compared to a passively-heated disk. %This should most significantly impact the inner disk, where JWST is most sensitive. 
The location and degree to which the disk is impacted by accretion heating will be reflected in JWST spectra and is expected to directly impact the observable mass of molecules. Before trends between stellar properties, disk substructure, or environment and MIR spectra can be robustly interpreted, it is critical to first understand how accretion rate and accretion luminosity shape the observable molecular reservoir.

%The fact that MIR line luminosity increases with observed accretion rate then can be simply assumed to attributed to an increased emitting radius and temperature. 

%However, the location of viscous disk accretion could influence the relative impact on line fluxes and ratios, thus the location of accretion heating within the disk can be constrained with JWST observations. Additionally, in Romero-Mirza et al. 2025 sub., it has been shown that the accretion luminosity - more so than the stellar luminosity - is correlated with observable water flux. This suggests that the correlation between disk accretion rate and water flux may be more than just a simple increase in the central temperature. Thermo-chemical models that take into account accretion heating can be used to put constraints on the accretion mechanisms within protoplanetary disks. 

In this article, we seek to model the impact of viscous accretion heating and accretion luminosity the observability of the mid-IR water lines. %We choose to use water as our representative molecule to study how its snowline and observable mass changes with accretion rate. 
Water and its spatial distribution in a disk is a vital parameter for planet-formation models, since its snowline is thought to enhance dust coagulation and growth of planetesimals, in addition to being a critical ingredient to life \citep{Jang-Condell04, Ros13,Drazkowska17}.
The Disk-Exoplanet C/Onnection (DECO) ALMA Large Program and JWST program (PI: Ilse Cleeves; 2022.1.00875.L; PI: Ilse Cleeves, ID: 3228) provides a sample of protoplanetary disks representative of the most common types of disks around stars, those around K and M-types. 
\cite{CarlosThesis} provides a uniform analysis of water lines from the 40 total sources within the JWST-DECO survey, providing observable water masses for 20 disks each in the Lupus and Taurus star forming regions. Half of the sample is classified as ``compact'' by having a millimeter dust radius $<$ 40 au and the other half is ``extended.'' We utilize a thermo-chemical model which implements accretion heating within the disk, and determine the resultant 2D chemical distribution of water. 

This paper is structured as follows. We elaborate on our experimental design in Section \ref{sec:Methods}. The impact that both viscous midplane heating and the illuminated contribution of L$_{acc}$ have on the temperature structure, resultant water snowline, and observable water reservoir are discussed in Section \ref{sec:Results}. The water snowline definition, the trend of observed water flux and L$_{acc}$ from \citep{CarlosThesis} is reproduced and discussed, and the critical component of accretion heating is discussed in Section \ref{sec:Discussion}, and finally we summarize our findings in Section \ref{sec:Conclusion}.

%the correlation between accretion luminosity and the observed gaseous reservoir in disks as probed with JWST. We Romero-Mirza et al. 2026 sub. (hereafter RM25)

%We elaborate on our experimental design in Section \ref{sec:Methods}, discuss the implication of adding accretion heating and luminosity on the 2D snowline and observed spectra in Section \ref{sec:Results}, and discuss the implication of adding accretion at the midplane of the disk in Section \ref{sec:Discussion}. We summarize our take-away points in Section \ref{sec:Conclusion}.

\section{Methods} \label{sec:Methods}

This project utilizes the thermo-chemical code \textsc{DALI} \citep{Bruderer12,Bruderer13}. DALI takes in a user-defined physical structure of a disk, a stellar spectrum, chemical network and initial conditions. In \cite{Calahan25} an accretion module simulating viscous heating originating from the midplane was added. It follows work by \cite{DAlessio98, Lodato08, Harsono15} and depends on one assumed accretion rate across the midplane, and the physical structure of the disk. In this work, we also introduce the addition of the accretion luminosity to the stellar spectrum. We utilize PySynphot \citep{pysynphot} to simulate a stellar spectrum given a stellar luminosity, mass, and radius. The accretion luminosity component is modeled as a 8,000~K blackbody added to the spectra with an increased emitting area based on the accretion rate. 

\begin{figure}
    \centering
    \includegraphics[width=.9\linewidth]{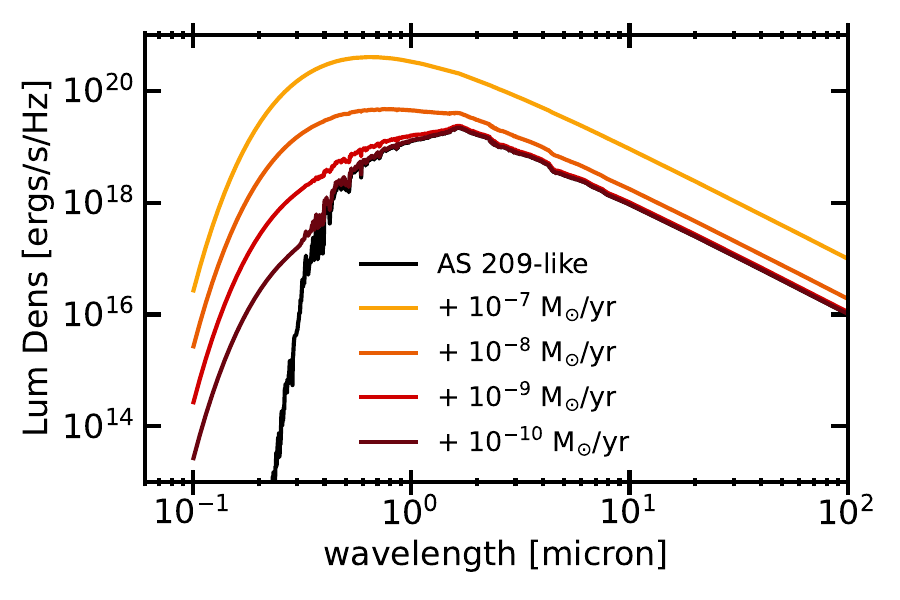}
    \caption{\textit{The stellar spectra of a star with the same stellar mass, luminosity, and effective temperature as AS 209 [black] and the addition of differing levels of accretion rate [brown through yellow]. }}
    \label{fig:stellar_spectra}
\end{figure}

We chose to use K-type stellar parameters with similar stellar parameters as AS~209, effective temperature of 4,040~K, bolometric luminosity of 1.4~L$_{\odot}$, and mass of 0.6~M$_{\odot}$\citep{Herbig88,Andrews18, Oberg21}. AS~209 is one of the more well studied K-stars, and representative system for both available observations to compare against, and solar system-like origins. The stellar spectrum plus the addition of accretion luminosity is shown in Figure \ref{fig:stellar_spectra}. We explore the accretion rates 10$^{-10}$, 10$^{-9}$, 10$^{-8}$, and 10$^{-7}$ M$_{\odot}$/yr. In \cite{Manara23}, accretion rates are listed for known protoplanetary disks at the time, and 10$^{-8}$ M$_{\odot}$/yr represents an average rate. We chose to explore multiple accretion rates below this average, as these are also found around Class II protoplanetary disks, and one step above, as anything beyond 10$^{-7}$ M$_{\odot}$/yr would be in line with an active protostar or outbursting source \citep{Hartmann96}. 

The stellar spectrum is the primary heating source for the disk and is used to calculate the internal radiation field and dust temperature in the first stage of calculations. Viscous heating is also added at this dust radiative transfer stage. After the dust temperature and radiation field are calculated, the gas temperature is determined, initialized by the dust temperature, and iterated based on the chemical initial conditions and network (as described further in \cite{Calahan25}, which is based on the UMIST 2006 network \citep{Woodall07}). 

For this project we seek to explore how adding accretion to the disk impacts the observable component of JWST observations. We start with a fiducial model that is representative of a typical disk around a K-star. Our fiducial model has a stellar spectrum typical of a 0.6 M$_{\odot}$ star, and a disk that is 1\% of the stellar mass, with a total dust-to-gas ratio of $10^{-2}$. These values are typical to what has been observed and modeled for similar sources \citep[i.e. ,][]{Bosman22a,Vlasblom24}. We set up a user-defined physical structure following these parametrized equations for density \cite{Lynden-Bell74, Hartmann98, Andrews09,Du14} :

\begin{equation}
    \rho(r,z) = \frac{\Sigma}{\sqrt{2\pi}h}\rm{exp}\left[-\frac{1}{2}\left(\frac{z}{h}\right)^{2}\right],
\end{equation}

where,

\begin{equation}
    h=h_{c}\left(\frac{r}{r_{c}}\right)^{\psi}
\end{equation}

and 

\begin{equation}
    \Sigma(r)=\Sigma_{c}\left(\frac{r}{r_{c}}\right)^{-\gamma} \rm{exp}\left[-(\frac{r}{r_{c}})^{2-\gamma}\right].
\end{equation}

\noindent We solve for $\Sigma_{c}$ by defining a total disk mass, and outer and inner disk radius:

\begin{equation}
    \Sigma_{c} = M_{\rm{gas}} (\frac{2-\gamma}{2\pi r_{c}^2})[\exp^{-(r_{in}/r_{c})^{2-\gamma}}-\exp^{-(r_{out}/r_{c})^{2-\gamma}}]
\end{equation}

\noindent Our disk is smooth, with a moderate level of flaring and critical heights and radius; see Table \ref{tab:struct_properties}.

\begin{deluxetable}{llr}
\vspace{.1cm}
\label{tab:struct_properties}
\tablecolumns{7}
\tablewidth{0pt}
\tabletypesize{\small}
\tablecaption{Disk Model Parameters}
\tablehead{Factor	& Value & Description }
 \startdata
 M$_{\rm{gas}}$	&	 0.006 M$_{\odot}$ & Total disk gas mass	\\
M$_{\rm{dust}}$/M$_{\rm{gas}}$	&	10$^{-2}$ &	Gas to dust mass ratio	\\
M$_{\rm{small\ dust}}$	&	M$_{\rm{dust}}$ $\times$ 0.01&	Total mass of the small dust population	\\
M$_{\rm{large\ dust}}$	&	M$_{\rm{dust}}$ $\times$ 0.99&	Total mass of the large dust population		\\
r$_{in}$	&	0.1 AU& Inner radius of the disk \& grid		\\
r$_{out}$	&	100 AU&	Outer radius of the disk \& grid	\\	
r$_{c}$	&	10 AU&	Critical radius; gas and dust populations		\\
h$_{c}$ [gas and small dust]	&	1 AU&	Critical height; applicable for 	\\
	&	&	small dust and gas populations	\\
h$_{c}$ [large grains]	&	0.1 AU & Critical height; only applicable		\\
	&	 &  for large dust population		\\
$\gamma$	&	1.0 &	Power index for disk surface density	\\
$\psi$	&	1.05 &	 Power index for scale height		\\
$\dot{\rm{M}}$	&	10$^{-10}$, 10$^{-9}$, & Accretion rate\\
&10$^{-8}$, 10$^{-7}$ M$_{\odot}$/yr &		\\
 \enddata
\tablecomments{User defined physical structure was calculated using equations 1-4 and the above values.}
\end{deluxetable}

\section{Results} \label{sec:Results}

The midplane viscous accretion heating, and the addition of $L_{\rm{acc}}$ to the stellar spectrum both increase the temperature of the disk with increasing accretion rate. The contribution of accretion heating from these two mechanisms, however, contribute to the heating in different ways. We isolate the impact of viscous heating and  L$_{\rm{acc}}$ and describe in detail where in the disk the excess heating is primarily deposited. We then go on to explore different accretion rates, and how the temperature and snowline are impacted.

\subsection{The Impact on the 2D Gas Temperature Profile}

\subsubsection{Isolating the Impact from Viscous Midplane Heating and L$_{\rm{acc}}$}

The addition of viscous midplane accretion and excess UV-rich luminosity from accretion on to the star have markedly different impacts on the resultant disk temperature structure.  \cite{D'Alessio98} explored the impact of these two heat sources (in addition to others such as cosmic ray and radioactive heating), in a numerical model that could reproduce the Spectral Energy Distribution (SED) of classical T Tauri stars. They find in their typical T Tauri disk model that stellar heating is the primary heat source beyond 1-2 au, and that stellar irradiation increases the temperature beyond that of a viscous non-irradiated disk. This was also was found in \cite{Calahan25} with a thermo-chemical modeling tool, with the impact of viscous heating being primarily located towards the midplane of the inner disk. 

\begin{figure*}
    \centering
    \includegraphics[width=.9\linewidth]{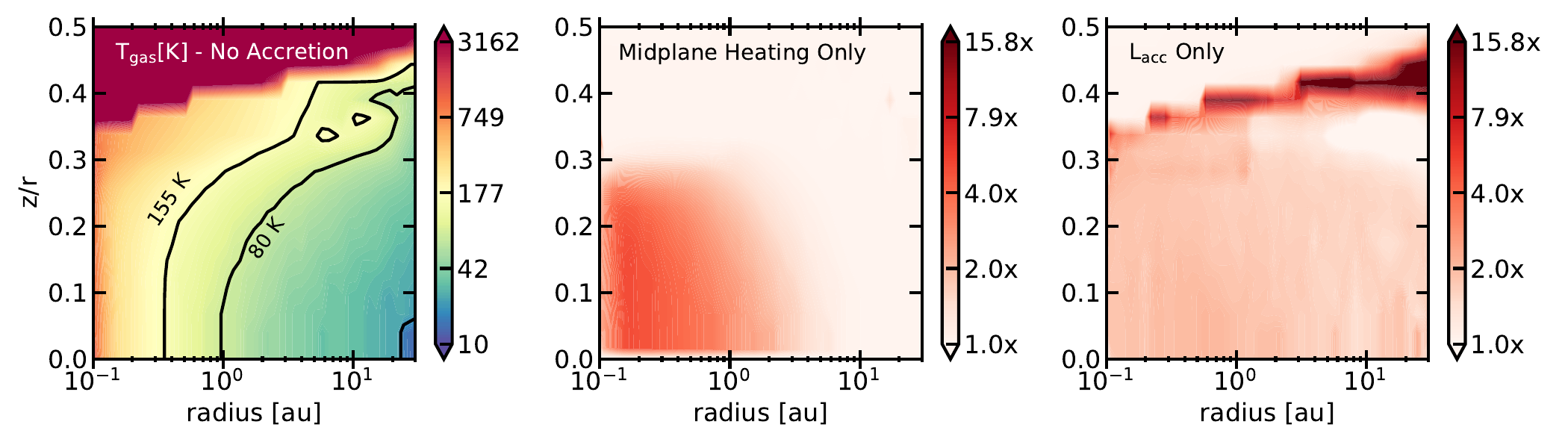}
    \caption{\textit{Left: The gas temperature profile of a model with no accretion, and contour lines at 155 and 80~K. Middle: The factor representing the increase in gas temperature with the addition of viscous heating originating at the midplane with 10$^{-8}$ M$_{\odot}$/yr . Right: The factor representing the increase in gas temperature due to the accretion luminosity component to the stellar spectrum.}}
    \label{fig:wheres_heat}
\end{figure*}

In this study, we isolate the impact of viscous accretion and the addition of L$_{acc}$ within the context of the 2D inner disk structure. Our fiducial model shows an increased temperature from midplane-localized heating being confined to within 1~au (see Fig. \ref{fig:wheres_heat}). The increase in temperature continues up to a z/r$\approx$0.3 within this physical structure, although this specific limit will be dependent on the distribution of the vertical column density and gas-to-dust ratio. A `flatter', more vertically compact disk, with more mass concentrated towards the midplane will not contribute as much heat to the upper layer of the disk (z/r$>\sim$0.3). The addition of an 8,000~K blackbody to the stellar spectra has the strongest impact on the temperature in the upper layers of the disk, and farther out than where viscous midplane heating is prevalent (see Fig. \ref{fig:wheres_heat}). 

%\begin{figure*}
%    \centering
%    \includegraphics[width=.9\linewidth]{figures/temperature_slices.pdf}
%    \caption{\textit{Left: The temperature profile of our fiducial model with vertical dashed lines at 0.9~au, 2.1~au, and 4.7~au, corresponding to the right-most plots. The three plots on the right are vertical slices of temperature, with each line corresponding to the temperature in the fiducial model, one with added L$_{\rm{acc}}$ to the stellar spectrum, one with just the midplane addition, and the dashed line with both added.}}
%    \label{fig:temperatures}
%\end{figure*}

Once combined the viscous heating and L$_{\rm{acc}}$ contribute to the final temperature structure at different relative amounts based on radius. The further out into the disk, the more the L$_{\rm{acc}}$ contribution is the primary driver of the increased temperature. 
In our fiducial model, the midplane temperature at 1~au starts at $\approx$70~K. With the addition of L$_{\rm{acc}}$ corresponding to an accretion rate of 10$^{-8}$ M$_{\odot}$/yr, the temperature at 1~au is $\approx$150~K, and combined with viscous heating in the midplane, it jumps to $>200$~K. At around 2~au in the midplane, the contribution from L$_{\rm{acc}}$ and viscous accretion are comparable, contributing equally to increase the final temperature once combined, and at 4~au, L$_{\rm{acc}}$ is the primary driver of any increase in temperature compared to a non-accreting disk. The accretion rate and physical structure will all affect these specific quoted values, however it is clear that these forms of accretion heating will impact various molecular snow surfaces and the observable spatial extent of the disk of which JWST is sensitive. %The midplane is also heated by the added central luminosity; there is an approximately a twofold increase in temperature in our fiducial $\dot{M}=10^{-8}$ M$_{\odot}$/yr model within 10 au. 

%This is shown in Figure \ref{fig:temperatures}. 

\subsubsection{The Relative Impact of Differing Accretion Rates}

\begin{figure*}
    \centering
    \includegraphics[width=1\linewidth]{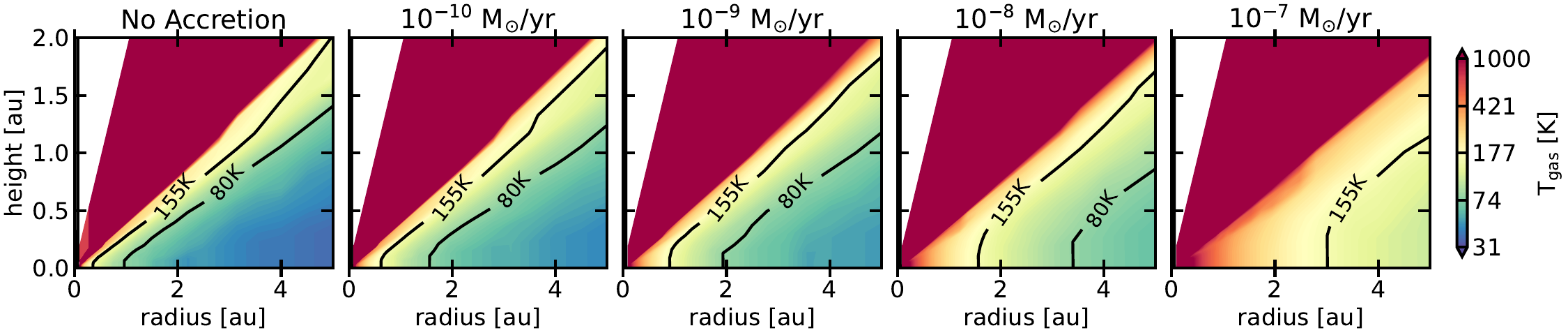}
    \caption{\textit{2D temperature plots across different accretion rates, with both viscous heating and L$_{acc}$ implemented. The temperature contours of 155~K and 80~K approximately trace the H$_{2}$O and CO$_{2}$ sublimation fronts, respectively. All plots are shown with the same color-scale.}}
    \label{fig:nice_temps}
\end{figure*}

\begin{figure}
    \centering
    \includegraphics[width=1.1\linewidth]{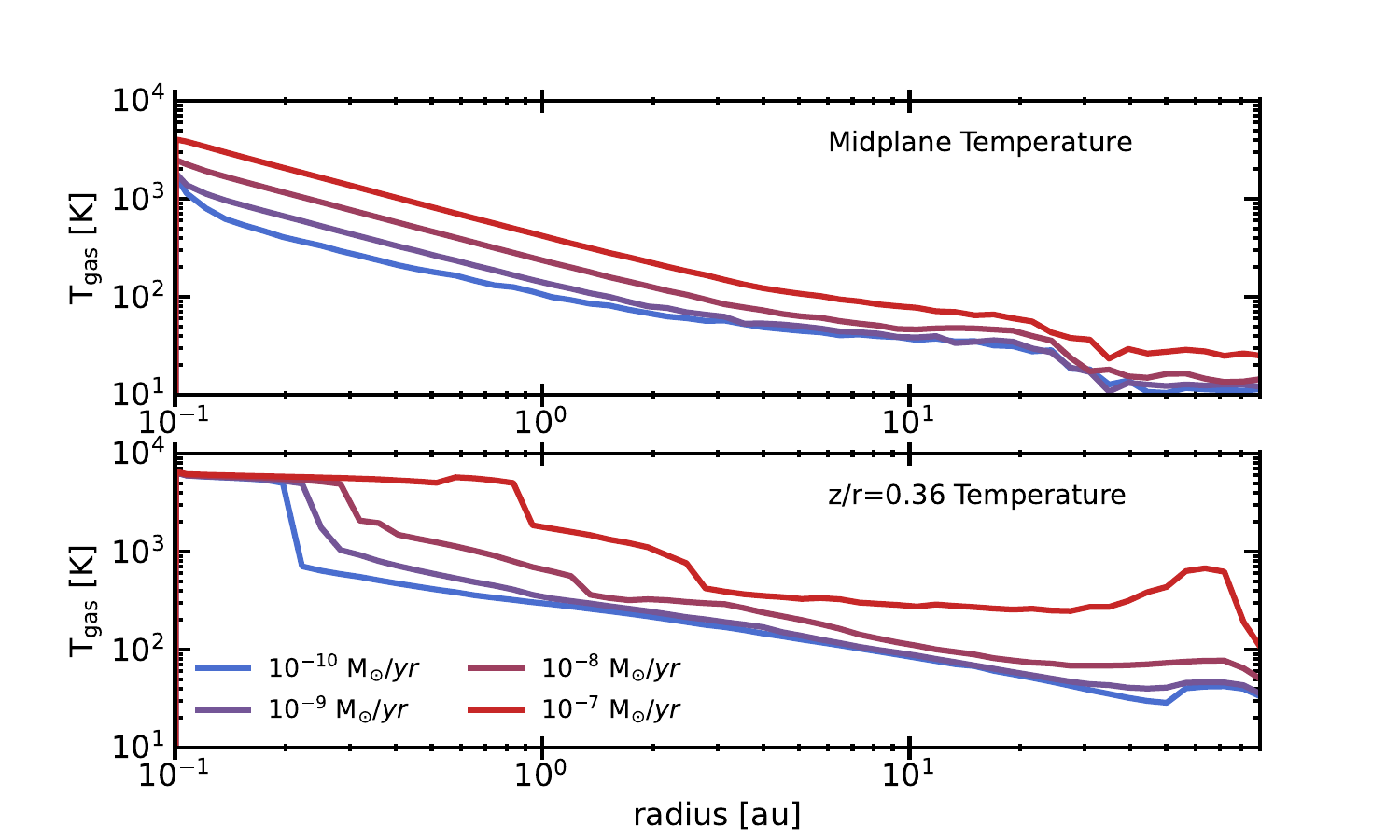}
    \caption{\textit{The radial temperature of the disk across the midplane (top) and z/r=0.36 (bottom), with each line corresponding to a different accretion rate. }}
    \label{fig:midplane_atmos}
\end{figure}

With increasing accretion rate, the temperature increases (see Fig. \ref{fig:nice_temps}), most significantly in the inner disk and atmosphere. The 155~K and 80~K isotherms both move outwards with higher accretion rate, and the distance between the two increases. At the highest accretion rate, the gas temperature within 3~au of the star is all above 155~K. The radial location at which the gas temperature diverges from a passively heated disk increases with accretion rate; extending to 100~au at 10$^{-7}$ M$_{\odot}$/yr. At the midplane and within 1~au, the temperature increases uniformly in log-space with each order of magnitude increase in accretion rate (see Fig. \ref{fig:midplane_atmos}). 

Higher up in the disk, the impact on gas temperature from accretion can be more extreme. In Figure \ref{fig:midplane_atmos} we also show the change in temperature with accretion rate at $\approx$3 gas pressure scale heights in the disk, or z/r=0.36 in this physical structure. At this height, the transition between atomic-dominated and molecular-dominated gas occurs at a radius of $\approx$0.2~au for non-accreting/low accreting disks. This transition is clearly seen when the gas temperature goes from 10,000~K to $\sim$1000s of K, as the presence of molecules and PAHs start to contribute to cooling of the system. The location of this transition is pushed farther back in radius with increasing accretion rate, at first gradually (always within 0.4~au at a z/r=0.36), but then a more extreme radial distance with the highest accretion rate (now at 1 au at 10$^{-7}$ M$_{\odot}$/yr), atoms dominate the gas component z/r$\gtrapprox$0.3. %At this height in the atmosphere, the temperatures of a disk with a 10$^{-10}$ and 10$^{-9}$ M$_{\odot}$/yr come to an agreement at around 2-3~au, while the 10$^{-8}$ M$_{\odot}$/yr takes until beyond 10~au, and  10$^{-7}$ M$_{\odot}$/yr doesn't converge with the low accretion temperature until beyond 100~au. 

The change in disk temperature due to differing accretion rates impact the full disk, with a more gradual increase in temperature at the midplane as compared to the atmosphere. Both the L$_{\rm{acc}}$ and viscous midplane heating impact the final temperature in the midplane, but with each order of magnitude step in accretion, the temperature increase is uniform in log space, and pushes the spatial extent of the accretion-heating dominated region of the disk to larger radii. Thus, a higher accretion rate does not uniformally heat up the disk, but will have a varying degree of impact both vertically and radially. 

\subsection{Snow Surfaces Depend on Temperature and Density}

The snow surface of a molecule is the sublimation front, where a molecule transitions from gas-phase to ice-phase. A molecule's snow surface, and its intersection at the midplane (snowline), can influence the gas/ice phase elemental ratios, and can impact the material composition of dust grains, which strongly influence the efficiency of planet formation\citep{Jang-Condell04, Ros13}. Water is the most abundant volatile species frozen-out onto dust grains in disk, and its snow surface has been shown to alter different aspects of planet formation \citep[e.g.][]{Ciesla06,Cridland16,Drazkowska17}. The water snow surface not only impacts the gas/ice-phase water reservoir; once water sublimates dust grains are effectively stripped-bare, and any other volatiles trapped within a water matrix can escape as well \citep{Simon23, Williams25}. The water snow surface shapes planetary formation, final exoplanet elemental ratios, and the abundances of volatiles in the gas-phase, thus its location must be understood. Water freezes out at approximately 120-170~K \citep{Lodders03}, with 155~K being a common approximation. This corresponds to a midplane snowline at $\approx$1~au in a disk around a K-star, thus we would expect the snow surface location should be strongly impacted by the addition of accretion heating.

\begin{figure}
    \centering
    \includegraphics[width=.6\linewidth]{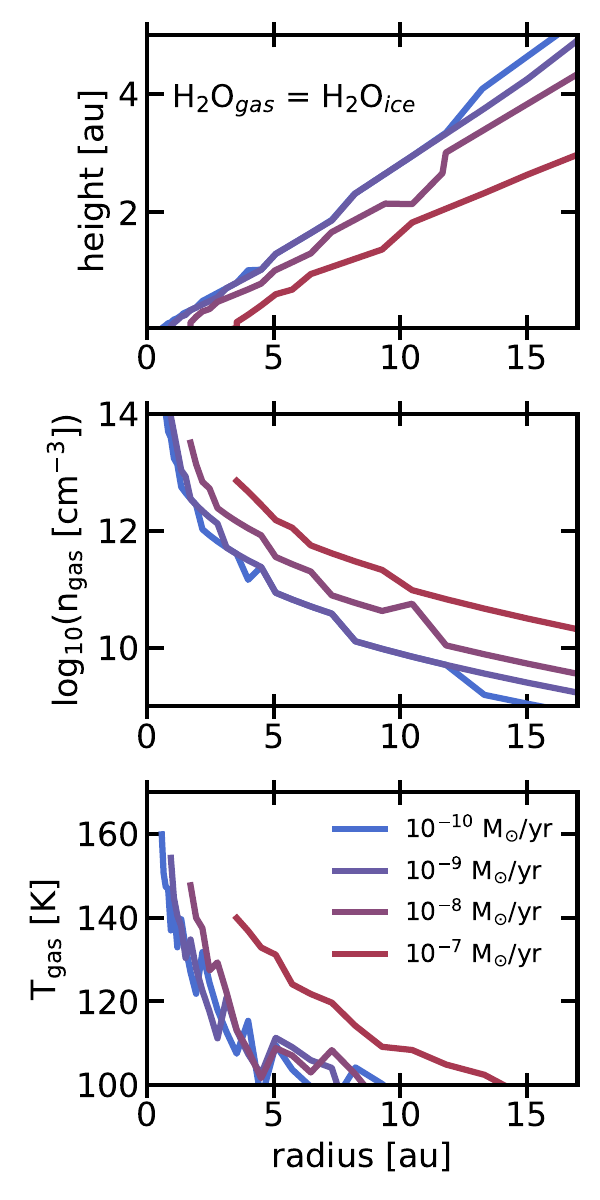}
    \caption{\textit{Top: The location of the water snowline, as defined by where the gas-phase H$_{2}$O abundances is equal to the ice-phase H$_{2}$O. Middle: The associated gas density along each of the water snow surfaces for each accretion rate. Bottom: The temperature along each water snowline for each accretion rate. The snowline intersects the midplane at different temperatures for each accretion rate. From lowest to highest accretion rate, the midplane snowline corresponds to 153, 149, 142, and 139~K. }}
    \label{fig:snowlines}
\end{figure}

We explore the impact that higher accretion rate has on snow surface location. To do so, we define snow surface as where there are equal amounts of ice-phase and gas-phase water. With higher accretion rate, the water midplane snowline is pushed to larger radii; originally being located within 1~au in our no accretion/low accretion models, and located at 3~au in our highest accretion model. Figure \ref{fig:snowlines} shows the water snow surface, as defined to be where the abundance of gas-phase water and ice-phase water are equal to each other. We also show the gas density and temperature along each water snow surface for different accretion rates. We see that not only does the radial location change, but the gas density and temperature at which water freeze out changes with accretion rate. At 10$^{-10}$ M$_{\odot}$/yr, the water snow surface intersects at the midplane at a temperature near 153~K. With each order of magnitude step in accretion rate, the midplane snowline is located at 149, 142, and finally at 139~K for a disk with 10$^{-7}$ M$_{\odot}$/yr. This decrease in the temperature at which water freezes out at corresponds to an decrease in gas density, and can be more clearly seen while tracing the vertical extent of the snow surface. This interplay among gas density, temperature, and molecular sticking properties should all be accounted for when determining the freeze-out temperature of a molecule \citep{Lodders03,Hollenbach09}, and is discussed further in Section 4.1. %This deviation from the assumed water freeze-out temperature can be seen as we trace the snowline up towards the atmosphere. As the density decreases with increasing height, so does the temperature at which there is an equal amount of gas and ice-phase water. 

\subsection{Accretion's Impact on Observable Water}

\begin{figure*}
    \centering
    \includegraphics[width=1\linewidth]{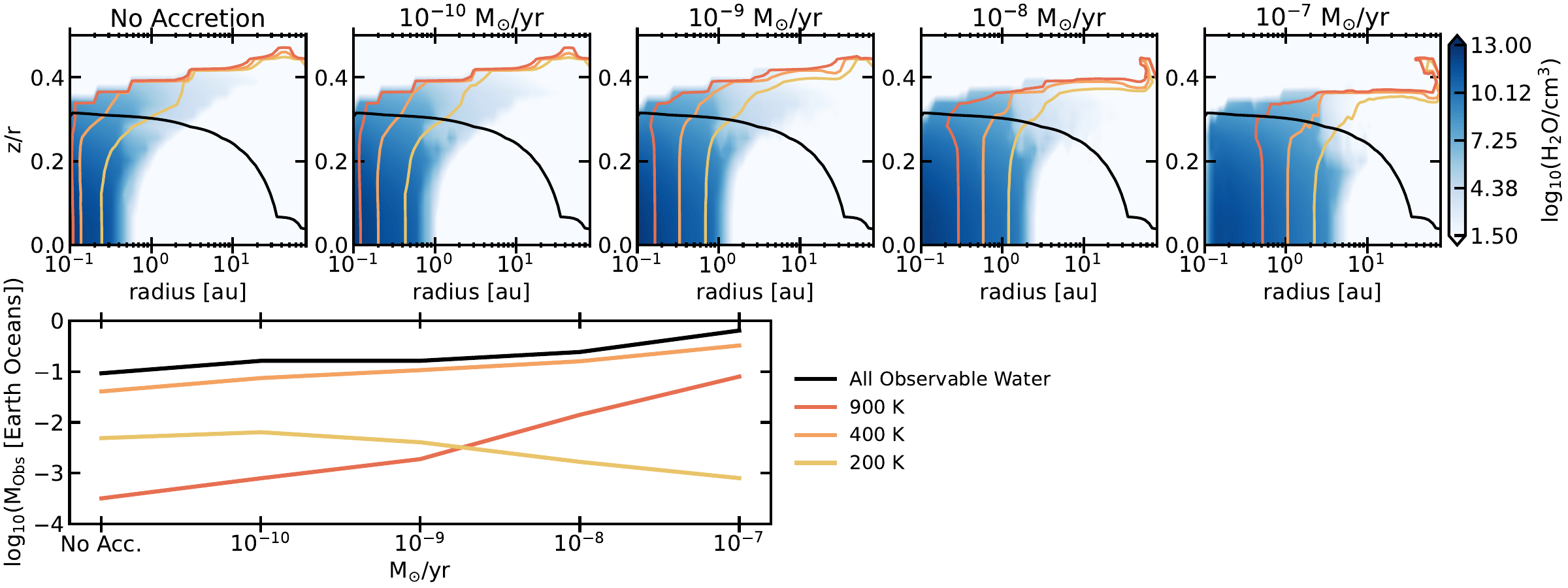}
    \caption{\textit{2D number densities of gas-phase water throughout the disk (top). The yellow, orange, and red contours correspond to 200, 400, and 900~K, delineating the cool, warm, and hot reservoirs respectively. The black contour follows where the disk is optically thick to IR wavelengths (10 $\mu$m). The bottom plot shows the total mass of water surrounding each of these contours $\pm$20\% in units of Earth's Oceans, and the total possible observable mass above the optically thick surface of the dust (black).}}
    \label{fig:2D_water_abu}
\end{figure*}

The water abundance within the inner few au is the available reservoir for terrestrial planet formation. %What is observed with JWST is likely just a small percentage of the full amount, as much of it will be hidden under optically thick dust and water itself becomes optically thick fairly high in the disk atmosphere under typical assumptions\citep{Woitke18, Calahan22a}. 
In this analysis, we calculate the observable water mass and determine how this population changes with accretion rate. We define the observable water mass as any water above the region in which dust become optically thick to infrared wavelengths. This definition on its own may lead to an overestimate of the observable water reservoir, as once H$_{2}^{16}$O becomes optically thick, it can obscure the otherwise observable H$_{2}^{16}$O below. However, in our fiducial case the optically thick IR boundary coincides closely with the point at which water emission also becomes optically thick. This apparent convergence is due to our assumptions in the dust distribution. Gas-to-dust ratio (100), percent of mass in the small grain population (1\%), scale height and radius (see Table \ref{tab:struct_properties}) all can change where the dust's optically thick surface resides. In a model with a gas-to-dust ratio of 1000, for example, the dust surface would be lower than the water $\tau$=1 layer.  We also separate the water reservoir into `cool' ($\approx$200~K), `warm' ($\approx$400~K), and `hot' ($\approx$900~K) constituents, as motivated by previous slab models of water spectra \citep{RomeroMirza24,CarlosThesis,Vlasblom25,Temmink25}. To calculate the observable mass at each temperature, we sum the water density within regions of the disk that are $\pm$25\% of 200, 400, and 900~K. Figure \ref{fig:2D_water_abu} shows the water number density in our fiducial model, for each accretion rate we explore and the summed total water mass within $\pm$25\% of each temperature of interest. The thermal contours separating the different temperature water reservoirs are pushed to larger radii with increasing accretion rate, as seen in previous sections, and the optically thick surface to IR does not change as the physical structure remains the same. The boundaries of the observable water reservoirs change with accretion rate both radially and vertically.% At high accretion rate, $>10^{-8}$ M$_{\odot}$/yr, there is a region within 0.1~au where the inner disk is so hot it becomes atomic-rich, and water dissociates into OH and atomic O. 
The emissive water is distributed over larger radii with increasing accretion rate.  %Figure \ref{fig:water_over_acc} shows how the observable water mass for each temperature population increases with accretion rate. 

There is an increase in observable water mass due to the increase in emitting area. The hot water reservoir increases at a more rapid pace than the warm and cool populations. The hot and cool populations seem to converge or even flip in relative observable mass with increasing accretion rate, while the warm reservoir ($\sim$400~K) consistently probes the most water mass (see Fig. \ref{fig:2D_water_abu}). The different dependencies of each water reservoir on the accretion luminosity can be understood through their evolving emitting regions. %, the isothermal contours not only are pushed farther out, but also pushed a bit deeper into the disk, closer to the optically thick boundary.
There is still an increase in the total abundance of gas-phase water with higher accretion luminosity, as we define these reservoirs, but there is a limit to how much of the water mass is still observable due to the location of the dust optically thick surface. For the cold population (gas at 200$\pm$25\% K), we see a decrease in observable water mass with accretion rate. This is due not only to the cold water that is now being produced and hidden underneath the dust optically thick surface, but also to the stronger gradient of temperature; less of the spatial extent of the disk is within 200$\pm$25\% K. Additionally, in the outer atmosphere, the increase in L$_{acc}$ also increases the UV flux, photodissociating cool water in the outer disk atmosphere.  Many other physical mechanisms and properties of the disk can impact the observable water mass \citep{Houge25}, but all disks have some assumed accretion rate. Now that we have established a framework for interpreting how accretion rate impacts the observed water reservoir, it will be easier to disentangle other critical factors of a protoplanetary disk system. 

%\begin{figure}
%    \centering
%    \includegraphics[width=.9\linewidth]{figures/water_masses.pdf}
%    \caption{\textit{The observable water mass for the cool, warm, and hot water reservoirs at differing accretion rates seen within our models.}}
%    \label{fig:water_over_acc}
%\end{figure}

\section{Discussion} \label{sec:Discussion}

The presented models reproduce one of the strongest trends in observed JWST spectra: the increase of observed water luminosity/mass with stellar accretion rate. %In our implementation of accretion, the inner disk is more strongly impacted by the addition of accretion heating, thus water and other molecules with snowlines within the inner few au will be most impacted.
Here, we discuss the implications of our results, how it may impact planet formation theory and how our models compare to JWST slab modeling and constraints on the observed water mass.  

\subsection{Implications for the Water Snow Surface}

\begin{figure}
    \centering
    \includegraphics[width=1\linewidth]{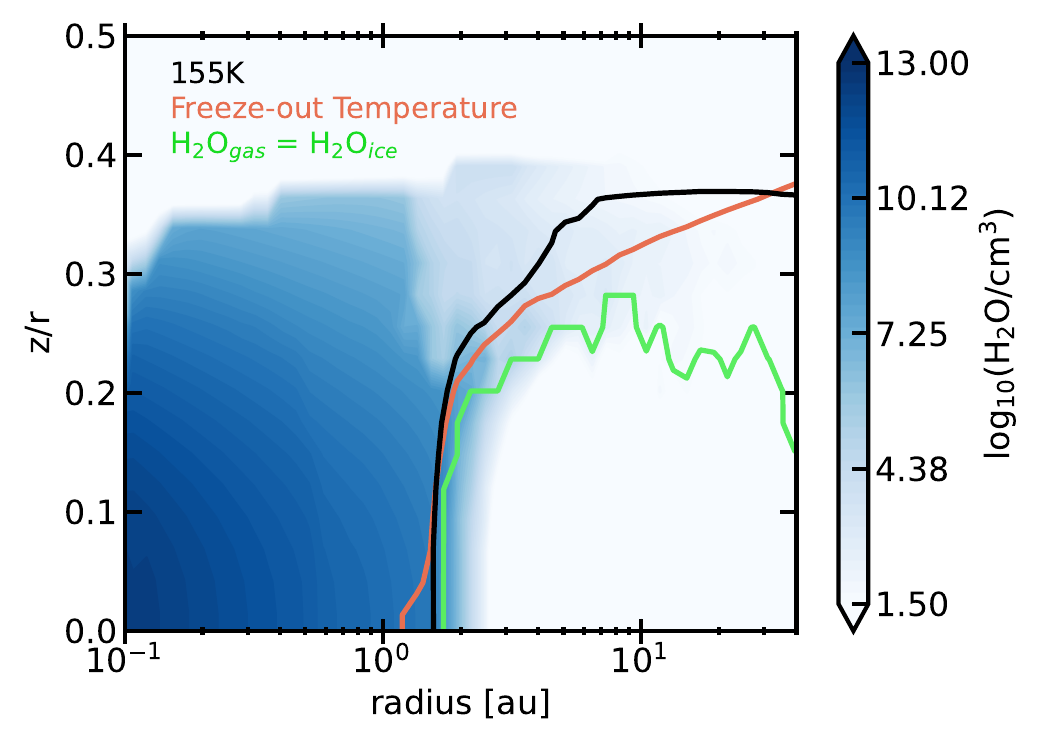}
    \caption{\textit{The gas-phase water number density in our model with an accretion rate of 10$^{-8}$ M$_{\odot}$/yr, with different definitions of the snow surface. Black corresponds to the isothermal contour of 155~K, orange to the theoretical freeze-out temperature (see Equation 1), and blue is where the abundance of gas-phase and ice-phase water are equal.}}
    \label{fig:differing_snowlines}
\end{figure}

The snow surface of a given molecule can be defined in different ways depending on a given science goal. The sublimation front of a molecule can be defined as: (1) following an isothermal contour along the temperature at the midplane where a molecule freezes out \citep[i.e.][]{Hayashi81, Oberg11}, (2) the theoretical temperature at which a molecule will freeze-out based on molecular properties and gas density assuming desorption is dominated by thermal processes\citep[i.e.][]{Lodders03}, or (3) as where there are an equal number of molecules in the gas and ice-phase taking into account the combined effect of chemistry, radiative environment, and thermal structure in 2D \citep[as has been explored in the context of water in ][]{Leemker21}. We compare each of these ways to predict the sublimation front in Figure \ref{fig:differing_snowlines} using water as an example. The theoretical boundary (definition no. 2) at which 50\% of the water should be frozen out by setting $n_{ice}=n_{gas}$ is set by the following equation,

\begin{equation}
    \frac{n_{ice}}{n_{gas}} = \frac{n_{dust} \pi a_{dust}^{2}(3k_{B}T_{gas}/m_{X})^{1/2}}{\nu_{1} \exp(-E_{b}/T_{dust})},
\end{equation}

\noindent which comes from \cite{Harsono15} and references therein. We determine where this boundary is in our model by using $T_{gas}$ and $T_{dust}$ from our thermo-chemical model, and $n_{dust}$ from our physical setup. $a_{dust}$ is an effective grain size, which we set as constant at 0.1$\micron$, $k_{B}$ is the boltzmann constant, $m_{X}$ is the mass of the species (in our case water), and $E_{b}$ is the binding energy \citep[5773~K,][]{Fraser01}. $\nu_{1}$ is the first-order pre-exponential factor, and is defined as follows \citep{Hasegawa92,Walsh10}:

\begin{equation}
    \nu_{1} = \sqrt\frac{2N_{\rm{ss}}E_{b}k_{B}}{\pi^{2}m_{\rm{X}}} \rm{s}^{-1}
\end{equation}

\noindent Here, $N_{\rm{ss}}$ is the number of binding sites per surface area \citep[8$\times10^{14}$ cm$^{-2}$,][]{Visser11}.

These different definitions of snow surface converge at the midplane (see Fig \ref{fig:differing_snowlines}), but diverge at larger radii. The 155~K isocontour is located above the other snow surface boundaries by about half a scale height, and thus could be considered an over estimate of the water sublimation front. The difference between the 155~K isocontour and the theoretical freeze-out temperature derives from the changing density. The boundary where the gas- and ice-phase water abundances are equal, as calculated by our thermo-chemical code, is even lower than this theoretical limit, by about another half of a scale height. These layers are strongly affected by chemical processing. In this small region, oxygen is not primarily stored in water, but rather in gas-phase atomic O, CO$_{2}$, and increased but trace amounts of O$_{2}$. The spatial difference between the theoretical water-ice line and the thermo-chemical ice line increases with radius and could have trickle down effects, impacting water delivery and planet formation efficiency. 

The water snow surface definitions diverge above a z/r=0.2 in our fiducial model. However, in this region there is little gas-phase water, it is instead the location at which the ice-phase water resides which defines the thermo-chemically calculated snow surface. The abundance of ice/gas phase water around the snow surface can impact the water and its isotopologue ratios inherited onto forming planets, and these models suggest beyond a few au, the water is primarily in the ice-phase, even a 2 scale heights and above. However, these are results from a static chemical model, in a dynamic model it could be that water ice can fill in this other-wise water-poor region between the theoretical 50\% water ice boundary and the thermo-chemically calculated 50\% water ice boundary \citep{Krijt16}. If adding dynamics to this framework replenished gas-phase water above the snow surface, it can mix downwards, replenishing the ice-phase water reservoir \citep[e.g.][]{Vlasblom25}.  Obtaining observational constraints on this region and the abundance of gas/ice-phase water is probed best in the far-IR \citep[as has been done in ][]{Hogerheijde11,Du2017} and will be a target of upcoming missions (i.e. The PRobe Far-Infrared Mission for Astrophysics, PRIMA, or the Planetary Origins and Evolution Multispectral Monochromator, POEMM). 

\subsection{A Comparison to JWST Observations}

The positive correlation seen in observed fluxes/molecular mass and accretion luminosity seems straightforward. The increase in central luminosity will heat up the disk, pushing out isothermal lines to larger radii, hence increasing the overall emission surface. We found this to be true, but due to both observational limits and the change in the gradient of the temperature profile, the cold, warm, and hot reservoirs are impacted in relatively distinct ways with increased accretion rate. This agrees well with observations, first discovered with the correlation of accretion luminosity and water line luminosity as a function of upper level energy \citep{Banzatti23,Banzatti25}, and then interpreted into observable water mass of different temperature reservoirs \citep{CarlosThesis}.

\begin{figure*}
    \centering
    \includegraphics[width=1\linewidth]{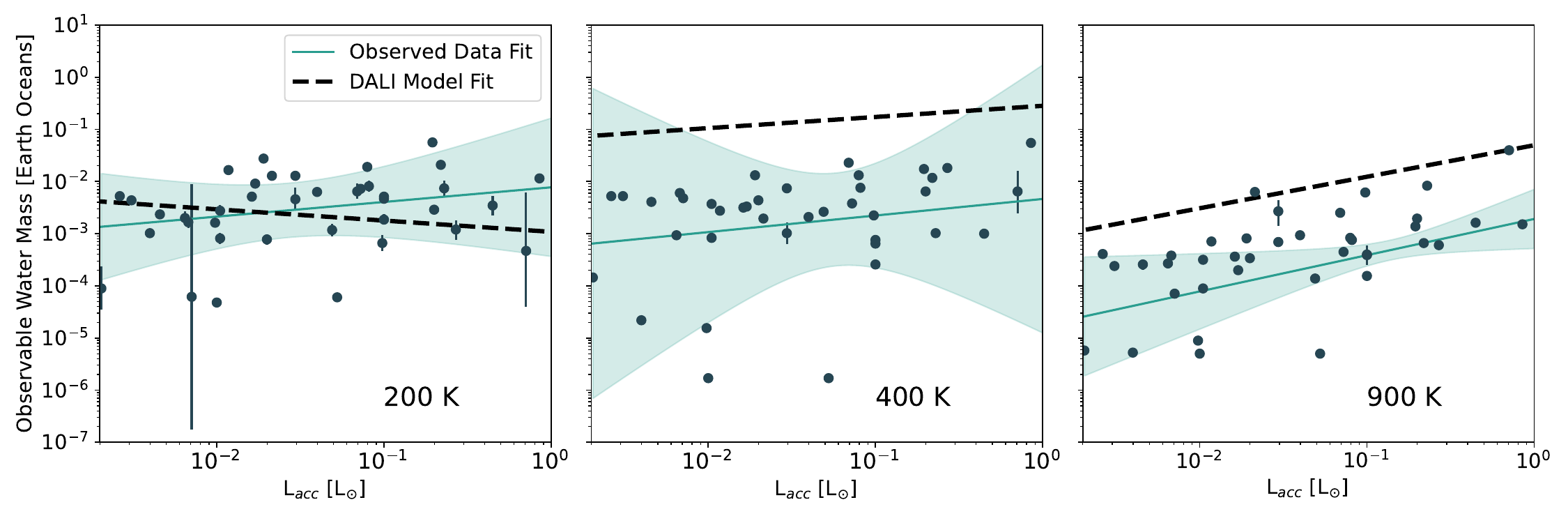}
    \caption{\textit{Observable water mass found in disks around M and K stars in Lupus and Taurus, as calculated from \cite{CarlosThesis}, and their corresponding L$_{acc}$ [L$_{\odot}$]. Fits to the data with 95\% confidence intervals are in teal. Dashed line is the observable water mass from the DALI K-star model in from this work for the cold, warm, and hot populations.}}
    \label{fig:compare_2_obs}
\end{figure*}

In Figure \ref{fig:compare_2_obs}, we compare our observed water mass vs accretion rate to those found in \cite{CarlosThesis}. We choose compare observed mass as derived by a slab model and from the thermo-chemical code for ease of comparison, and it has been shown that slab modeling of thermo-chemical simulated spectra produce similar results \citep{Vlasblom25}. The fiducial model from \citep{Vlasblom25} and our ``No Accretion'' model have similar distributions of water and temperature. Once accretion is added, due to both midplane accretion and the added luminosity from accretion there are minimal temperature inversion in the disk, localized only to the inner 0.3~AU, at temperatures higher than was we use to estimate the `hot' reservoir of water. Thus we move forward with comparing our thermo-chemical observed water mass and retrievals from slab models. To compare to the calculated observable mass from slab modeling, we add up the total water mass $\pm$25\% of the 200, 400, and 900~K isothermal contours. 
We fit both the observed data points and the DALI model output from the five models with different accretion rates to a powerlaw. The exponents to the powerlaw of each thermal water reservoir for the DALI output agrees within 95\% confidence range to the data (see Table \ref{tab:expon}). In our cold water reservoir, there is a negative correlation between observed water mass and accretion luminosity. The spatial extent of the cold water reservoir not only is pushed below the optically thick surface, but the gradient of temperature radially becomes more extreme, with an increasingly smaller volume of the disk having gas temperatures within 25\% of 200K. Additionally there is a decrease in gas-phase water formation in the outer atmosphere of the disk due to an increase in UV flux with higher accretion rate. 

Our fiducial model also roughly reproduces the observed water mass seen in the JWST-DECO sample, lying within the uncertainty of the observations or within an order of magnitude of them. The observed masses derived from our model should be treated as an upper limit. We total up the total water mass down to the optically thick dust surface, but there are many factors that could limit the amount of water the observations are sensitive to. Primarily, the water lines used to derive the observed populations in \citep{CarlosThesis} are optically thick \citep{Woitke24, Vlasblom25}, and are primarily sensitive to a region of the disk above the optically thick surface. The location of the dust optically thick surface can be pushed up, namely if the small grains are more abundant in the atmosphere than our model assumed. We explored one disk mass and one stellar type, 6$\times$10$^{-3}$ M$_{\odot}$ and a K star spectra. The disk mass is representative of an average disk mass in the sample, and the 0.6 M$_{\odot}$, 1.5 L$_{\odot}$ star is on the warmer end of the stellar hosts in the JWST-DECO survey. With a lower mass star, the water snowline will be located closer to the star, resulting in a smaller emitting area, equating to a lower water flux. 

%Our model results tend to lay above or at the highest end of the data, however the disk masses within the survey range over two orders of magnitude, with 6$\times$10$^{-3}$ M$_{\odot}$ being the highest mass and the same mass as our fiducial model. }

\begin{deluxetable}{lcc}
\vspace{.1cm}
\label{tab:expon}
\tablecolumns{3}
\tablewidth{0pt}
\tabletypesize{\small}
\tablecaption{Exponent Fits}
\tablehead{ Temperature Component	&	$b_{\rm{DALI}}$ 	&	$b_{\rm{data}}$ 	\\ }
 \startdata
200K	&	-0.22	&	0.28 $\pm$ 0.78	\\
400K	&	0.21	&	0.32 $\pm$ 1.9	\\
900K	&	0.60	&	0.69 $\pm$ 0.62	\\
\enddata
\tablecomments{The following functional form was fit to the observed data and DALI results: $M_{obs}=a\times L_{acc}^{b}$}
\end{deluxetable}

While we reproduce the critical trend of differing degrees of correlation with observable water mass and accretion rate, there are parameters of the physical structure that can change our results to a more favorable fit, even in this static case. The observational limit on what we consider `observable' water will strongly impact our results. We determine water to be in the observable extent of the disk if it is above an optically thick dust layer to Mid-IR wavelengths. The treatment of our dust population will impact where this limit is. If we were to use a higher gas-to-dust mass ratio, our optically thick surface would be pushed downwards into the disk, exposing more water (see discussion in Appendix). The observable water mass will be higher in our model than in the data, and each population will have a different power-law coefficient for the relation between observable mass and accretion luminosity. However, with the most common assumptions on disk physical structure for a disk around a K-type star, the fact that one DALI setup can reproduce trends seen in the data suggest that our results are robust, and can be used as a tool to interpret JWST spectra and constrain the observable water mass. 

\subsection{Does Midplane Heating Impact JWST Spectra?}

\begin{figure}
    \centering
    \includegraphics[width=1\linewidth]{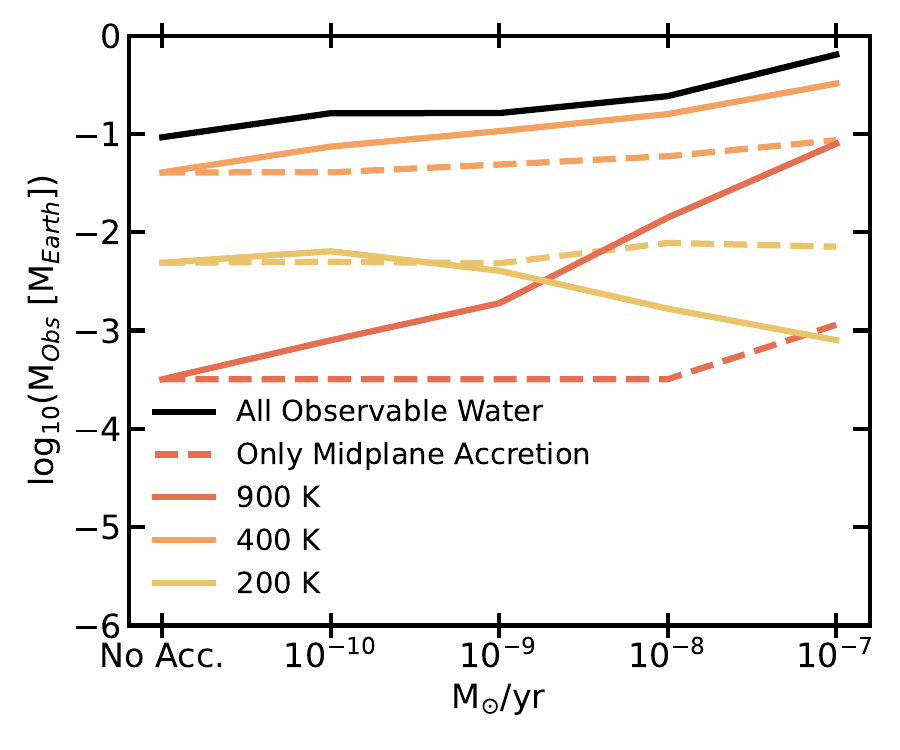}
    \caption{\textit{The observed water mass (in Earth's Oceans) versus accretion rate in the DALI models, with the dashed line showing the evolution with only heating due to midplane accretion, and the solid line as our final results with both midplane and illuminated contribution from L$_{acc}$. }}
    \label{fig:only_midplane}
\end{figure}

In this study, we implemented accretion heating onto a disk through both viscous heating in the midplane and an addition of a 8,000~K blackbody to the stellar spectrum. 8,000~K is not universally used as the associated temperaure due to accretion heating onto the star, however is commonly used, and we find no impact on our results due to the exact temperature value used, rather it is the total central luminosity that alters the temperature and water abundance. However, since the observed water reservoir is constrained to the disk atmosphere, it is the central radiative heating that has the strongest impact on our results. Viscous heating in the midplane pushed out the sublimation front of the molecule, but does not impact what is directly observable with JWST. If we were to remove the 8,000~K blackbody contribution, and only include midplane heating, we do not reproduce the observed trend in water mass and accretion rate (see Figure \ref{fig:only_midplane}).  

Viscous heating from disk accretion does not only necessarily originate from the midplane, however, it can also be centralized in the region of the disk that is has an active magnetorotational instability (MRI). This could be co-located with the atmosphere of the disk, where the observable water reservoir resides \citep[i.e. ,][]{Bai11,Perez-becker11}. We are able to reproduce observed trends with water mass without the addition of viscous accretion in the atmosphere, but this is worth studying in the future, as such an exercise could constrain the impact of MRI-activated accretion heating. 

\section{Conclusion} \label{sec:Conclusion}

In this article we explore how accounting for accretion heating impacts the thermal structure of a thermo-chemical model, and the resultant impact on the observed water reservoir. We find the following:

\begin{itemize}
    \item Midplane viscous accretion heating only impacts the temperature of the disk within the inner few au. The contribution of accretion to the central stellar luminosity has a stronger impact on the disk overall, especially in the atmosphere. 

    \item Molecular snow surfaces will be pushed outward to larger radii with increasing accretion rate. The freeze-out temperature at the midplane is close to commonly assumed 155~K for a no accretion case, but can decrease by 20~K for disks associated with higher accretion rates. 

    \item With the addition of accretion, our thermo-chemical models reproduce one of the strongest trends in IR molecular spectroscopy of T Tauri disks, the increase of observable water luminosity/mass with accretion luminosity. We also reproduce the decreasing strength of this trend in hot, warm, and cool water, respectively.

    %\item The total mass of the observable water reservoir in our thermo-chemical model is in agreement with that derived from slab model and JWST water spectra. It does not appear that we are severely missing or are overestimating the typical water mass observed in planet-forming disks.

    \item The trend in increasing observable mass and accretion rate is primarily due to the contribution of L$_{acc}$, and midplane heating is not required to reproduce observed trends. 
\end{itemize}

Mid-IR observations offer us a direct look at only a small percentage of the full population of planetary building blocks. It is critical to understand the region that we can directly observe, in order to ultimately gain clarity on the critical planet-forming mechanism occurring beneath this surface. We find that the contribution of accretion heating on the disk thermal and chemical structure can be understood through  a thermo-chemical model. The original hypothesis of increasing emitting area with higher accretion rate is confirmed \citep{Salyk11, Banzatti20, Vlasblom25}, with subtle details becoming clear only after considering the full dust and gas structure along with the thermal profile and chemical distribution simultaneously. 

%% Please use the acknowledgment and contribution environments. This will 
%% be anonomyized when the "anonymous" style option is used. 
\begin{acknowledgments}
We want to thank Jonathan McDowell for his 30+ years of dedication to running the Smithsonian Astrophysical Observatory's Research Experience for Undergraduates supported by the NSF, and Center for Astrophysics. 
J.K.C. acknowledges the Kavli-Laukien Origins of Life Fellowship,
K.I.\"O. acknowledges Simons Foundation grant
No. 686302, and an award from the Simons Foundation (grant No. 321183FY19). 
Support for C.J.L. was provided by NASA through the NASA Hubble Fellowship grant No. HST-HF2-51535.001-A awarded by the Space Telescope Science Institute, which is operated by the Association of Universities for Research in Astronomy, Inc., for NASA, under contract NAS5-26555. 
F.A. is funded by the European Union (ERC, UNVEIL, 101076613). Views and opinions expressed are however those of the author(s) only and do not necessarily reflect those of the European Union or the European Research Council. Neither the European Union nor the granting authority can be held responsible for them. 
T.M. was supported by the Royal Society, award numbers URF\textbackslash R1\textbackslash 211799 and RF\textbackslash ERE\textbackslash 231082. 
L.I.C. acknowledges support from the Research Corporation for Science Advancement Cottrell Scholarship Award 28249, the David and Lucille Packard Foundation, and NSF AAG awards 2205698 and 2407547, and JWST-GO-03228.001-A.
SK and TK acknowledge support from STFC Grant ST/Y002415/1. C.H.-V. and 
C.G-R. acknowledge support from ANID -- Millennium Science Initiative Program -- Center Code NCN2024\_001.
AG acknowledge support from NSF AST-2407547 and the David and Lucile Packard Foundation and the Virginia Institute of Theoretical Astronomy (VITA). 
JW is funded by the UK Science and Technology Facilities Council (STFC), grant code ST/Y509383/1
This work is based on observations made with the NASA/ESA/CSA James Webb Space Telescope. The data were obtained from the Mikulski Archive for Space Telescopes at the Space Telescope Science Institute, which is operated by the Association of Universities for Research in Astronomy, Inc., under NASA contract NAS 5-03127 for JWST. These observations are associated with programs \# 1282, 1584, 1640, 3228 \dataset[10.17909/w88w-w896]{http://dx.doi.org/10.17909/w88w-w896}.

\end{acknowledgments}

%\begin{contribution}
%%This section gives authors the space to recognize author contributions. The text inside this environment is NOT counted towards the total word quanta. At a minimum, manuscripts are expected to include this text:

%JC was the primary advisor for TD and was the primary author of the text, figures, and final model runs

%TD was a summer student in 2024 and tested the initial addition of accretion heating to DALI, and identified the change in the associated snowline temperature with accretion rate. TD and JC iterated closely over the academic year of 2024-2025 to hone in on the project

%K\{"}O advised both JC and TD throughout this project

%\end{contribution}

\software{astropy \citep{astropy:2013, astropy:2018} , 
          numpy \citep{harris2020array},
          DALI \citep{Bruderer12, Bruderer13}, 
          }

\appendix

\setcounter{figure}{0}
\setcounter{table}{0}
\renewcommand{\thefigure}{\arabic{figure}A}
\renewcommand{\thetable}{\arabic{table}A}

\section{Results from a Disk Setup with Depleted Dust Population}

The driving trend to dampen the positive correlation of observed water mass and accretion rate is the location of the optically thick dust boundary. Any change in assumption of the dust opacity can impact where this boundary lies, including material property, surface density distribution, or total mass of the dust population. In this modeling framework, a key parameter is the gas-to-dust ratio. The gas mass is defined, and the dust mass is determined based on the assumed ratio. We assume a gas-to-dust ratio of 100 in our presented fiducial model, as that is a widely used assumption \citep{Armitage11,Williams11}. However, modeling efforts focused on reproducing trends in the IR have used a gas-to-dust ratio of 1000 as well \citep{Bosman22a, Calahan22a}, and we explore how our final results may change with that different assumption. 

\begin{figure}
    \centering
    \includegraphics[width=0.8\linewidth]{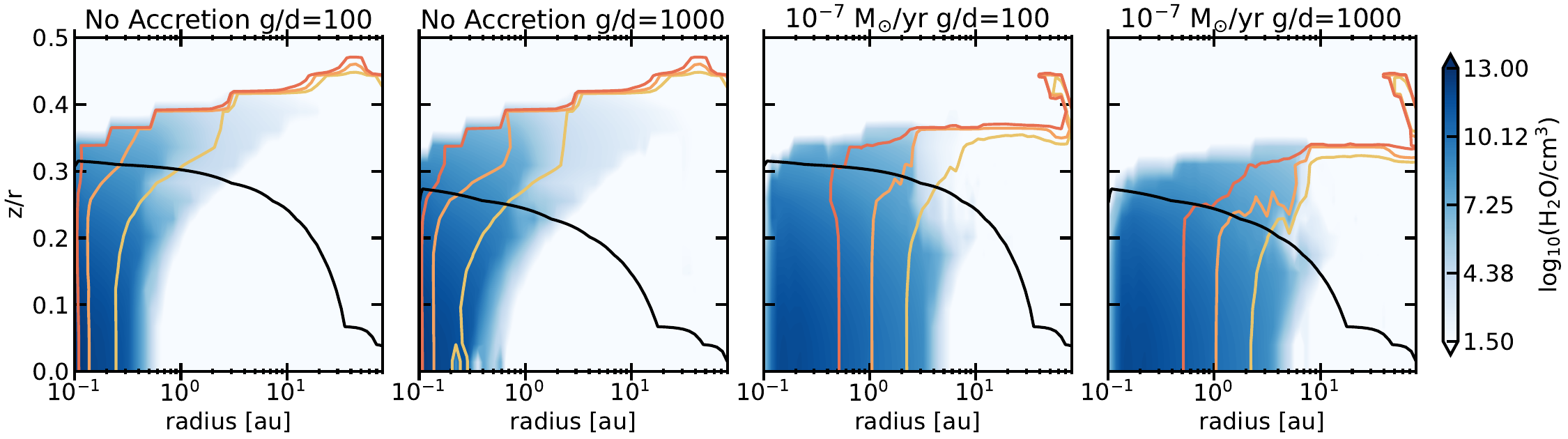}
    \caption{\textit{The gas-phase water number density in a model with no accretion (left two plots) and with a 10$^{-7}$ M$_{\odot}$/yr (right two plots) with different gas-to-dust (g/d) ratios. Red, orange, and yellow correspond to the 900, 400, and 200~K isotherms, and the black contour is where the disk is optically thick to the IR.}}
    \label{fig:100_vs_1000}
\end{figure}

\begin{figure}
    \centering
    \includegraphics[width=0.8\linewidth]{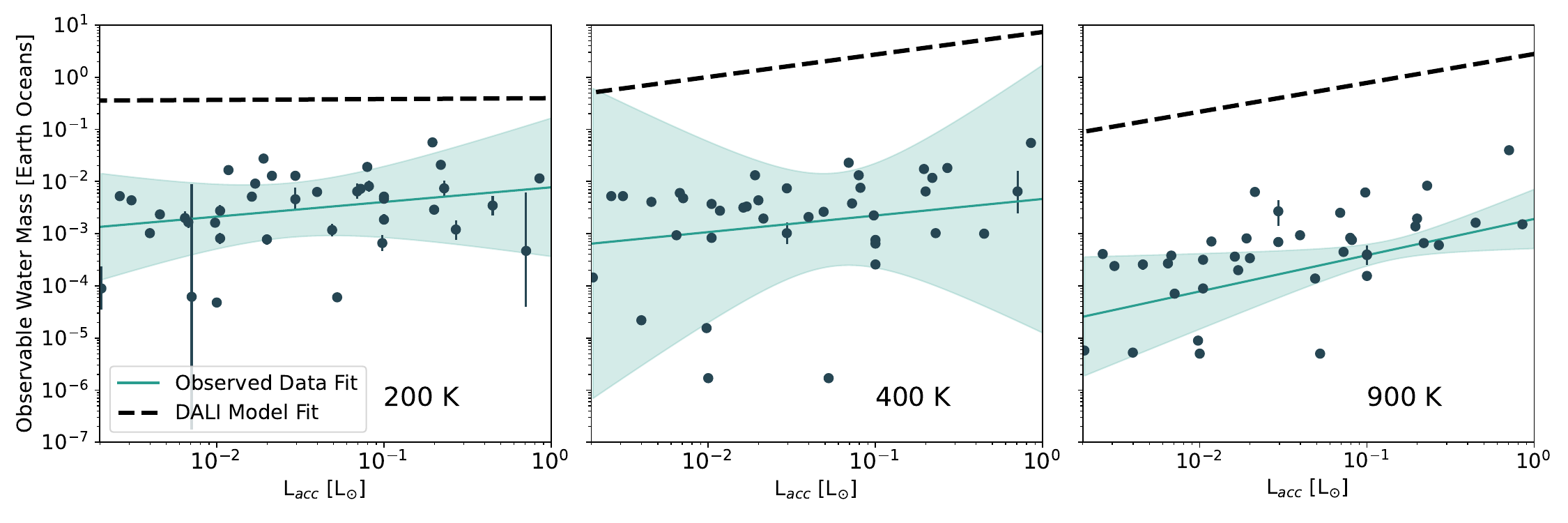}
    \caption{\textit{Observable water mass found in M and K stars in Lupus and Taurus, identical to Figure \ref{fig:compare_2_obs}, but with the DALI results from a model with a gas-to-dust ratio of 1000, instead of our fidiucial model with a gas-to-dust ratio of 100. }}
    \label{fig:compare_to_obs_gd1000}
\end{figure}

\begin{deluxetable}{lcc}
\vspace{.1cm}
\label{tab:expon2}
\tablecolumns{3}
\tablewidth{0pt}
\tabletypesize{\small}
\tablecaption{Exponent Fits}
\tablehead{ Temperature Component	&	$b_{\rm{DALI}}$ 	&	$b_{\rm{data}}$ 	\\ }
 \startdata
200K	&	0.08	&	0.28 $\pm$ 0.78	\\
400K	&	0.43	&	0.32 $\pm$ 1.9	\\
900K	&	0.55	&	0.69 $\pm$ 0.62	\\
\enddata
\tablecomments{The following functional form was fit to the observed data and DALI results for a model with a gas to dust ratio of 1000: $y=a\times L_{acc}[L_{\odot}]^{b}$}
\end{deluxetable}

The distribution of gas-phase water in a disk with a gas-to-dust ratio of 1000 is shown in Figure \ref{fig:100_vs_1000}, with the results from our fiducial model shown alongside. The increase in the gas-to-dust ratio lowers the vertical boundary at which the disk becomes optically thick in the IR, thus more water is exposed to be possibly observable. The temperature structure also changes with an increased gas-to-dust ratio, especially in the atmosphere of the disk. This is most noticeable in the high accretion example, the gradient of temperature in the disk atmosphere across radius becomes more extreme with higher gas-to-dust ratio. The final observable water mass is calculated the same way as the fiducial case, and the resultant masses in compared to the data is shown in \ref{fig:compare_to_obs_gd1000} with corresponding exponents in the DALI and observed data fit in Table \ref{tab:expon2}. We continue to reproduce the trend in increasing correlation with accretion luminosity with increase in water temperature, but the observed mass is at least an order of magnitude too high. Water will become optically thick above the dust optically thick surface in this model, and can bring the `observable water mass' as calculated in the thermo-chemical context closer to what is observed.

\bibliography{references}{}
\bibliographystyle{aasjournalv7}

\end{document}